\documentclass[11pt]{article} 
\usepackage{amssymb}
\usepackage{amsmath}
\usepackage{setspace}
\usepackage{amsthm}
\usepackage{epsfig,color,colordvi,wasysym}
\usepackage{graphicx}

\usepackage[top=1in,bottom=1in,left=1in,right=1in]{geometry}

\title{GREAT: GRaphlet Edge-based network AlignmenT}

\author{\textbf{Joseph Crawford and Tijana Milenkovi\'{c}$^{*}$}\\
Department of Computer Science and Engineering, Interdisciplinary Center for \\ Network Science and Applications, and ECK Institute for Global Health \\
University of Notre Dame\\
\noindent$^*$Corresponding Author (E-mail: tmilenko@nd.edu)}

\date{}
\begin{document}
\maketitle



\vspace{-0.6cm}

\noindent\textbf{Abstract.} Network alignment aims to find regions of
topological or functional similarities between networks. In
computational biology, it can be used to transfer biological knowledge
from a well-studied species to a poorly-studied species between
aligned network regions. Typically, existing network aligners first
compute similarities between nodes in different networks (via a node
cost function) and then aim to find a high-scoring alignment (node
mapping between the networks) with respect to ``node conservation'',
typically the total node cost function over all aligned nodes. Only
after an alignment is constructed, the existing methods evaluate its
quality with respect to an alternative measure, such as ``edge
conservation''. Thus, we recently aimed to directly optimize edge
conservation while constructing an alignment, which improved alignment
quality. Here, we approach a novel idea of maximizing \emph{both}
node and edge conservation, and we also approach this idea from a
novel perspective, by aligning optimally \emph{edges} between networks
first in order to improve node cost function needed to then align well
nodes between the networks. In the process, unlike the existing
measures of edge conservation that treat each conserved edge the same,
we favor conserved edges that are topologically similar over conserved
edges that are topologically dissimilar. We show that our novel
method, which we call \textbf{GR}aphlet \textbf{E}dge
\textbf{A}lignmen\textbf{T} (GREAT), improves upon state-of-the-art
methods that aim to optimize node conservation only or edge
conservation only.

\section{Background, motivation, and our contribution}

The goal of network (or graph) alignment in computational biology is
to find regions of topological or functional similarities between
networks of different species. (We note, however, that network
alignment has applications in many domains
\cite{li2009rimom,Narayanan2011}.) The more biological network data is
becoming available, the more importance the problem of network
alignment gains. This is because network alignment can be used, for
example, to transfer functional (e.g., aging-related
\cite{MilenkovicACMBCB2013,Faisal2014a,Faisal2014}) knowledge from
well annotated species to poorly studied ones between aligned network
regions.

There are two categories of network alignment methods: local network
alignment (LNA) and global network alignment (GNA). LNA focuses on
optimizing similarity between local (small) regions of different
networks, plus, it allows for a region in one network to be mapped
to multiple regions in another network. This way, LNA is generally
unable to find large conserved subgraphs between networks, and also,
it can lead to many-to-many node mappings between the networks
\cite{PathBlast,Sharan2005,Flannick2006,Mawish,Berg04,Liang2006a,Berg2006,Mina20014,AlignNemo},
which might be motivated biologically, but such alignments are hard to
characterize in terms of topological alignment quality
\cite{GRAAL,HGRAAL}. On the other hand, GNA aims to optimize global
(overall) similarity between different networks, and in general
(although some exceptions exist \cite{IsoRankN}), it results in
one-to-one (i.e., injective) node mapping
\cite{Narayanan2011,MilenkovicACMBCB2013,Faisal2014a,GRAAL,HGRAAL,IsoRankN,Singh2007,Flannick2008,Singh2008,GraphM,MIGRAAL,GHOST,NETAL,Guo2009,NATALIE,NATALIE2,MAGNA,NewSurvey2014}.
As such, GNA is able to find large conserved subgraphs, and it also
allows for quantifying both topological and biological quality of the
resulting alignments. In this study, we focus on one-to-one GNA due to
its recent popularity (and henceforth, we refer to GNA simply as
network alignment), but all concepts introduced here can be applied to
LNA as well.

We more formally define network alignment as an injective mapping
between the nodes of two networks that aligns the networks well with
respect to a desired criterion. Network alignment is a computationally
hard problem to solve due to the underlying subgraph isomorphism
problem, which is NP-complete \cite{West96}. The subgraph isomorphism
problem aims to find out whether some graph $G_2$ contains another
graph $G_1$ as its exact subgraph. With this in mind, the network
alignment problem aims to ``fit well'' $G_1$ into $G_2$ when $G_1$ is
not necessarily an exact subgraph of $G_2$. Thus, since network
alignment is computationally intractable, all existing algorithms aiming to solve this problem are heuristics.

In general (although there are some exceptions \cite{MAGNA}), existing
network alignment methods typically contain two algorithmic
components: 1) a node cost function and 2) an alignment strategy
\cite{MilenkovicACMBCB2013,Faisal2014a,GRAAL,HGRAAL,IsoRankN,Singh2007,MIGRAAL,GHOST,NETAL,Crawford2014,SPINAL}.
A node cost function finds pairwise topological (potentially also
biological, e.g., protein sequence) similarities (or equivalently,
costs) between two nodes from different networks, while the alignment
strategy uses these costs to select a high-scoring alignment (out of
all possible alignments) typically with respect to the total node cost
function over all aligned nodes \cite{MAGNA}. Then, the quality of the
resulting alignment is evaluated with respect to some other
topological measure, which is different than the node cost function
that is used to produce the alignment in the first place. (Alignment
quality is also measured via a biological measure, such as functional
enrichment of aligned node pairs \cite{MAGNA}.) Typically, one
measures the amount of conserved edges, and multiple measures have
been proposed for this purpose, with our recent symmetric substructure
score (S$^3$) being a superior measure \cite{MAGNA}. That is, the goal
of existing methods is to align nodes well in hope that they will
align edges well, but only \emph{after} the alignment is produced.
Hence, recently, we introduced a novel algorithm, called MAGNA, which
is capable of optimizing edge conservation directly \emph{while} an
alignment is being constructed \cite{MAGNA}. 

Here, we approach a novel idea of maximizing \emph{both} node and edge
conservation, and we also approach this idea from a novel perspective,
by aligning \emph{edges} between networks in order to improve node
cost function. These are the two major novelties of our study that
distinguish us from the existing work. In the process, unlike the
existing measures of edge conservation that treat each conserved edge
the same, we propose a new measure of edge conservation to favor
conserved edges that are topologically similar over conserved edges
that are topologically dissimilar. This is another of our novelties.
We note that a method exists that infers plausibly alignable
interactions across protein-protein interaction (PPI) networks of
different species \cite{Guo2009}. However, this method is guided
\emph{biologically} rather than topologically: it aligns PPIs relying
on conservation of their constituent domain interactions, and thus, it
aims to address not the problem of subgraph isomorphism nor edge
conservation but rather that of biological correctness of the aligned
edges \cite{Guo2009}.

To simultaneously optimize both node and edge conservation,
our new method, which we call GRaphlet Edge AlignmenT (GREAT), first
aims to optimally align edges between two networks, and based on the
resulting edge alignment, it constructs (as we will show) a more
efficient node cost function compared to state-of-the-art node
similarity measures that are typically used for this purpose. That is,
when we use within a given existing alignment strategy our new edge
alignment-based node cost function, we get alignments of higher
quality with respect to \emph{both} node and edge conservation than
when we use within the same alignment strategy an existing node cost
function. Thus, we improve upon methods that aim to optimize node
conservation only. At the same time, GREAT is comparable or superior
to MAGNA that aims to optimize edge conservation only.

\section{Methods}

GREAT consists of four algorithmic components: 1) edge cost function,
2) edge alignment strategy, 3) node cost function, and 4) node
alignment strategy. Edge cost function and edge alignment strategy are
used to align well edges between two networks, similar to how existing
methods align nodes between two networks based on node cost function
and node alignment strategy. Then, the resulting edge alignment is
used to compute a novel node cost function, according to which two
nodes from different networks are similar if the nodes' adjacent edges
have been aligned and with high similarity with respect to the edge
cost function. Then, the resulting edge alignment-based node cost
function is used within an existing node alignment strategy to produce
an injective node mapping between the networks. In this way, the
output of GREAT can be directly compared against alignments of the
existing methods. This section details the four steps of GREAT.

\subsection{GREAT's edge cost function and edge alignment strategy}\label{ealn}

To create pairwise edge scores, GREAT uses the notion of graphlets, as
follows.

\begin{figure}
\centering
\includegraphics[width=.685\columnwidth]{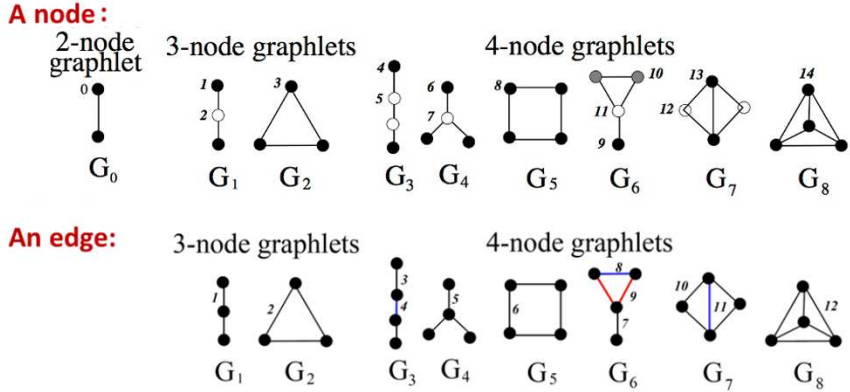}
\vspace{-0.5cm}
\caption{All automorphism orbits (i.e., topologically unique
``symmetry groups'') of a node (top) and an edge (bottom) in up to
4-node graphlets \cite{Przulj06ECCB,Solava2012}. We illustrate only
up to 4-node graphlets for esthetics, but all up to 5-node graphlets
(with 73 node orbits and 69 edge orbits) are used within our method.
The figure has been adopted and adapted from \cite{Solava2012}.}
\label{fig:graphlets}
\end{figure}

Graphlets are small induced non-isomorphic subgraphs (e.g., a triangle
or a square; Figure \ref{fig:graphlets})
\cite{Przulj04,Przulj06ECCB,Milenkovic2008,GraphCrunch,MMGP_Roy_Soc_09,Milenkovic2011,Solava2012,Hulovatyy2014}.
A graphlet-based \emph{node} cost function was already used for
network alignment by three state-of-the-art methods: GRAAL
\cite{GRAAL}, H-GRAAL \cite{HGRAAL}, and MI-GRAAL \cite{MIGRAAL}
(also, see \cite{MilenkovicACMBCB2013,Faisal2014a,Crawford2014}). This
node cost function relies on \emph{node graphlet degree vector}
(node-GDV) \cite{Przulj06ECCB}, which counts the number of graphlets
(i.e., their topologically unique node ``symmetry groups'', called
automorphism orbits; Figure \ref{fig:graphlets}) that a node touches.
Then, the graphlet-based node cost function computes topological
similarity between \emph{extended} neighborhoods of two nodes from
different networks by comparing the nodes' GDVs. Hence, this function
is called \emph{node-GDV-similarity} \cite{Milenkovic2008}. Recently,
we extended the notion of node-GDV into \emph{edge-GDV} (Figure
\ref{fig:graphlets}) and of node-GDV-similarity into
\emph{edge-GDV-similarity}, to allow for quantifying topological
similarity between extended neighborhoods of two \emph{edges} rather
than nodes \cite{Solava2012}. Then, we used edge-GDV-similarity as a
basis for a novel superior network clustering method
\cite{Solava2012}. Here, we use for the first time edge-GDV-similarity
for network alignment, and we use it as a part of our edge cost
function.

The other part of our edge cost function is a novel concept of
\emph{edge graphlet degree centrality} (edge-GDC), which we define as
a measure the complexity of the extended network neighborhood of an
edge (see below for a formal definition). We introduce edge-GDC to
modify the total similarity of aligning two edges, in order to favor
alignment of the densest parts of the networks. Namely, edges with a
given edge-GDV-similarity and with high edge-GDC (and thus dense
network neighborhoods) should be aligned before correspondingly
edge-GDV-similar edges with low edge-GDC \cite{HGRAAL}.

We define edge-GDC analogously to our existing definition of 
node-GDC \cite{Milenkovic2011}. For a given edge $e$, we denote the
$i^\text{th}$ coordinate of its edge-GDV (that is, the number of times
edge $e$ touches orbit $i$) as $e_i$. Then: $\text{edge-GDC}(e) =
\sum_1^{68}w_i * ln(e_i+1)$, where $w_i \in [0,1]$ is the weight of
orbit $i$ that accounts for dependencies between orbits, and 68 is the
total number of edge orbits in 3-5-node graphlets (there is an
additional orbit for the only 2-node graphlet, i.e., an edge, but we
leave out this orbit, as each edge will participate in exactly one
such orbit) \cite{Milenkovic2008,Solava2012}.

With the notions of edge-GDV-similarity and edge-GDC introduced, we
define our edge cost function (ECF), i.e., the total similarity
between two edges $e$ and $f$ from different networks, as: 

\vspace{-0.5cm}

\begin{equation} \label{egdcl}
\text{ECF} = \alpha \times \text{edge-GDV-similarity}(e,f) + (1-\alpha) \times \frac{\text{edge-GDC}(e) + \text{edge-GDC}(f)}{max(\text{edge-GDC($G_1$)}) + max(\text{edge-GDC($G_2$)})}
\end{equation}
where $\alpha$ is a parameter in [0,1], $G_1$ and $G_2$ are the two
networks being aligned, and edge-GDC($G_i$) is the maximum GDC in
network $G_i$. For this study, we vary $\alpha$ from 0 to 1 in
increments of 0.2. The formula 
is designed to normalize edge cost function to [0,1] range.

Given pairwise edge scores computed with respect to the above edge
cost function, GREAT feeds these scores into an existing edge
alignment strategy to produce injective edge mapping between the two
networks. We use two such strategies: 1) greedy alignment strategy,
which maps, one at a time, the highest-scoring edge pairs in a greedy
fashion, and 2) the Hungarian algorithm for maximum weight bipartite
matching, which produces optimal edge mapping with respect to our edge
cost function. We use these methods as edge alignment strategies
because equivalent (and thus comparable) methods have already been
used in the context of network alignment as node alignment strategies,
within e.g., IsoRank
\cite{Singh2007} and H-GRAAL \cite{HGRAAL}, respectively. Ideally, we would 
have adjusted more recent and superior node alignment strategies, such
as those of MI-GRAAL \cite{MIGRAAL} or GHOST \cite{GHOST}, to fit the
context of our edge alignment problem. However, generalizing these
node alignment strategies into analogous (and thus comparable) edge
alignment strategies is non-trivial, as the current implementations of
MI-GRAAL or GHOST either rely on proprietary libraries or are too
complex to be extended in any way, respectively.

\subsection{GREAT's node cost function and node alignment strategy}\label{NCF}

After generating an edge alignment, GREAT continues onto calculating
pairwise node scores based on this alignment, as follows. Let $v_1$ be a
node in graph $G_1$ and let $v_2$ be a node in graph $G_2$. Let $E'$ be the
set of aligned edges and let $sim(v_1,v_2)$ be the similarity between $v_1$
and $v_2$. Then, GREAT's edge alignment-based node cost function
$sim(v_1,v_2)$ is the sum of similarities (with respect to edge cost
function) over all edges in $E'$ (Appendix Figure \ref{ncf}).

After GREAT generates node cost function as above, it feeds the
resulting node scores into an existing node alignment strategy to
generate an injective node between the two. Here, we use three such
strategies: 1) greedy alignment strategy, 2) Hungarian algorithm for
maximum weight bipartite matching, and 3) MI-GRAAL's alignment
strategy. Again, we would have used a more recent node alignment
strategy, such as GHOST's, but the current implementation of GHOST is
too complex to allow for feeding into its alignment strategy any node
cost function but its own.

\subsection{Variations of GREAT}

By mixing and matching different options for GREAT's edge and node
alignment strategies, as discussed above, we get a total of six
different GREAT variations, i.e., aligners (Table~\ref{tab:great}).

\begin{table}[h]
\centering
\begin{tabular}{| l | l | l |}
\hline
GREAT variation & Edge alignment strategy & Node alignment strategy \\
\hline
EGG & Greedy & Greedy \\
\hline
EHG & Hungarian & Greedy \\
\hline
EGH& Greedy & Hungarian \\
\hline
EHH & Hungarian & Hungarian \\
\hline
EGM & Greedy & MI-GRAAL \\
\hline
EHM & Hungarian & MI-GRAAL \\
\hline
\end{tabular} 
\caption{Six variations of GREAT, depending on which edge and node alignment strategy it uses.}
\label{tab:great}
\end{table}

\subsection{Network data} \label{networks}

To test GREAT's performance, we use a popular evaluation test
\cite{GRAAL,HGRAAL,MIGRAAL,GHOST,MAGNA,MilenkovicACMBCB2013,Faisal2014a,Crawford2014}.
Namely, we focus on a high-confidence yeast PPI network with 1,004
proteins and 8,323 PPIs \cite{Collins07}, and we produce five
additional ``synthetic'' networks by adding noise to the yeast
network. The noise is the addition to the original yeast network of
$x\%$ of low-confidence PPIs from the same data set \cite{Collins07},
where we vary $x$ from 5\% to 25\% in increments of 5\%. We align the
original yeast network to each of the synthetic networks with $x\%$
noise, resulting in the total of five network pairs to be aligned.
Importantly, since all network pairs have the same set of nodes, we
know the true node correspondence (i.e., mapping). Thus, for each
considered method, we can measure how well the method reconstructs the
correspondence, along with evaluating the method's alignment quality
with respect to some other measures (Section
\ref{sect:methods_measures}).

We note the main focus of our paper is a fair evaluation of our new
GREAT method against the existing methods. If we aimed to predict new
biological knowledge, we would have applied our method to additional
networks, such as PPI networks of different species. However, since
our main focus is method evaluation, we focus on the above network
data set because: 1) the original yeast network is of high confidence
and thus trustable; 2) the data encompasses different PPI types,
including PPIs obtained via affinity purification followed by mass
spectrometry (AP/MS), and as such is of high coverage, 3) the same
data has already been actively used for evaluation of different
network aligners, and 4) we know the true node mapping as well as the
actual level of structural difference (corresponding to the given
percentage of the low-confidence PPIs) between each pair of aligned
networks, and hence, we can meaningfully interpret our alignments
(where this is not the case for networks with unknown node mapping or
unknown structural difference)
\cite{GRAAL,HGRAAL,MIGRAAL,GHOST,MilenkovicACMBCB2013,MAGNA,Faisal2014a}.
Ultimately, what matters for a fair evaluation is that all compared
methods are tested on the same data, which is exactly what we do
\cite{Faisal2014a}.

\subsection{Network alignment quality
measures}\label{sect:methods_measures}

Let $G_1=(V_1,E_1)$ and $G_2(V_2,E_2)$ be graphs such that $V_1 \leq
V_2$. Let $E_2'$ be the set of edges in $G_2$ that exist between the
set of nodes in $G_2$ that are aligned to nodes in $G_1$. Then, we
measure alignment quality with respect to the following
well-established measures
\cite{GRAAL,HGRAAL,MIGRAAL,GHOST,MAGNA,MilenkovicACMBCB2013,Faisal2014a,Crawford2014}.

\vspace{0.1cm}

\noindent\emph{Node correctness (NC):} If $f: V_1 \rightarrow V_2$ is
the correct node mapping
of $G_1$ to $G_2$ and $h: V_1 \rightarrow V_2$ is an alignment
produced by the given method, then $NC = \frac{|\{u \in V_1 : f(u) =
h(u)
\}|}{|V_1|}
\times 100\%$ \cite{GRAAL}. This measure can only be used on networks
with known node mapping, such as our data.

\vspace{0.1cm}

\noindent\emph{Symmetric substructure score (S$^3$)}: Although NC captures
the amount of true node mapping, it is still important to measure the
amount of conserved edges. For example, if we map an $n$-node clique
(complete graph) in one network to an $n$-node clique in another
network, there are many possible topologically correct alignments with
respect to S$^3$, i.e., alignments that conserve all edges, but there
is a single correct alignment with respect to NC. Plus, true node
mapping is not known for most real-world networks; in such cases, NC
can not be computed and one needs to rely on measures of edge
conservation. One such measure is S$^3$, defined as the percentage of
conserved edges out of all edges in $E_1$ and $E_2'$ combined. More
formally, it is defined as follows: $S^{3} = \frac{|E_{1} \cap
E'_2|}{|E_1| +|E'_2| - |E_{1} \cap E'_2|}\times{100\%}$
\cite{MAGNA}. Alternative measures of edge conservation exist, such as
edge correctness \cite{GRAAL} and induced conserved structure
\cite{GHOST}, but S$^{3}$ combines the advantages of both of these
measures while addressing their drawbacks, and as such, it has been
shown to be the superior of the three measures \cite{MAGNA}.

\vspace{0.1cm}

\noindent The size of the \emph{largest connected common subgraph
(LCCS)} \cite{GRAAL}, which we use because of two alignments with
similar S$^3$ scores, one could expose large, contiguous, and
topologically complex regions of network similarity, while the other
could fail to do so. Thus, in addition to counting aligned edges, it
is important that the aligned edges cluster together to form large
connected subgraphs rather than being isolated. Hence, a connected
common subgraph (CCS) is defined as a connected subgraph (not
necessarily induced) that appears in both networks \cite{HGRAAL}. We
measure the size of the largest CCS (LCCS) in terms of the number of
nodes as well as edges. Namely, we compute the LCCS score as in our
recent work \cite{MAGNA}. First, we count $N$, the percentage of nodes
from $G_1$ that are in the LCCS. Then, we count $E$, the percentage of
edges that are in the LCCS out of all edges that could have been
aligned between the nodes in the LCCS. That is, $E$ is the minimum of
the number of edges in the subgraph of $G_1$ that is induced on the
nodes from the LCCS, and the number of edges in the subgraph of $G_2$
that is induced on the nodes from the LCCS \cite{MAGNA}. Finally, we
compute their geometric mean as $\sqrt(N \times E)$, in order to
penalize alignments that have small $N$ or small $E$. Large values of
this final LCCS score are desirable.

\vspace{-0.3cm}

\section{Results and discussion}

We aim to answer the following: {\bf 1)} What parameter values to use
within GREAT's edge cost function to optimally balance between
edge-GDV-similarity and edge-GDC (Section \ref{GREAT_ECF_paramaters})?
{\bf 2)} Does edge-based network alignment improve upon comparable
traditional node-based network alignment (Section \ref{edge_vs_node})?
This is the main goal of our study, and achieving it would be
sufficient to demonstrate GREAT's superiority over the traditional
methods. {\bf 3)} Given that we will demonstrate that edge alignment
\emph{does} improve over node alignment, which of the two edge
alignment strategies (greedy versus optimal) to use to achieve
\emph{both} high accuracy and low computational complexity (Section
\ref{sec:time})? That is, can we still achieve with fast greedy edge
alignment similar accuracy as with slower optimal edge alignment? This
has important implications for processing large real-world networks.
{\bf 4)} Can GREAT, our edge-based network alignment method, not only
beat \emph{comparable} node-based network alignment methods (question
2 above) but also the \emph{most recent} and thus advanced existing
network aligners (which would only further confirm GREAT's
superiority) (Section \ref{GREAT_vs_existing})?

\vspace{-0.1cm}

\subsection{Best parameter values within GREAT's edge cost
function}\label{GREAT_ECF_paramaters}

\begin{figure}[h!]
\centering
\includegraphics[width=1\columnwidth]{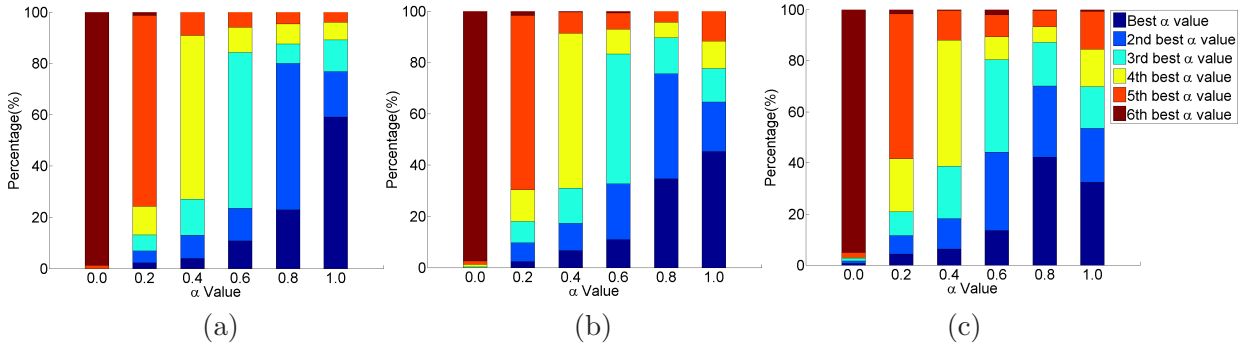}\\
\hspace{-.3in}(a) \hspace{1.7in}(b) \hspace{1.7in}(c) \\
\vspace{-0.2cm}
\caption{The ranking of the six $\alpha$ values used in GREAT's edge 
cost function across all variations of GREAT and all synthetic network
alignments with respect to: {\bf (a)} NC, {\bf (b)} S$^3$, and {\bf
(c)} LCCS. Recall that $\alpha=1$ corresponds to using only
edge-GDV-similarity, while $\alpha=0$ corresponds to using only
edge-GDC (Equation~\ref{egdcl}). }
\label{alpha}
\end{figure}

Recall from Section \ref{ealn} that the $\alpha$ parameter controls
the contribution of edge-GDV-similarity and edge-GDC in GREAT's edge
cost function. When we test its effect on a comprehensive network set
(see below), we find that overall, $\alpha$ of 1 is superior (although
$\alpha$ of 0.8 performs relatively well too) (Figure \ref{alpha}).
That is, edge-GDV-similarity is overall favored over edge-GDC. We note
that this was not the case with a comparable node cost function
\cite{HGRAAL}, where some contribution of node centrality was desired,
which is why we tested the effect of edge-GDC in GREAT's edge cost
function in the first place. Henceforth, in subsequent analyses, we
consider only the dominant $\alpha$ of 1.

We have performed this analysis on synthetic (geometric and scale-free
\cite{GraphCrunch}) networks of different sizes (we varied the number
of nodes from 500 to 1,000, and for a given node size, we varied the
average degree from four to 12). We have aligned each such network to
its noisy counterpart. Here, by noisy counterpart, we mean that in a
given synthetic network, we have rewired $x\%$ of the network's edges,
where $x\in\{5,12,15,20,25\}$. We have done this on the synthetic
(geometric and scale-free) network data rather than on our yeast
network data from Section~\ref{networks}, since we needed to test many
different values of $\alpha$, and doing so on relatively large (dense)
yeast networks is more time consuming. Also, we wanted to test the
effect of $\alpha$ on alignment quality as a function of network size,
and the yeast data does not allow for this. But henceforth, when we
test actual alignment accuracy, we focus only on the yeast network
data.

\vspace{-0.5cm}

\subsection{Is edge alignment worth it compared to node
alignment?}\label{edge_vs_node}

Recall that there are different GREAT variations depending on the
choices of edge and node alignment strategies (Table~\ref{tab:great}).
To fairly evaluate whether edge-based network alignment improves upon
node-based network alignment, we benchmark a given variation of GREAT
against the comparable node-based network alignment method. That is,
recall that a given version of GREAT uses edge-GDV-similarity-based
edge cost function (corresponding to $\alpha$ of 1), a given (greedy
or optimal) edge alignment strategy, the edge-alignment-based node
cost function, and a given (greedy, optimal (Hungarian), or
MI-GRAAL's) node alignment strategy. Given this, we produce the
corresponding node-based network alignment method as follows: it uses
node-GDV-similarity as its node cost function and the same node
alignment strategy as the given version of GREAT. This way, because
edge-GDV-similarity and node-GDV-similarity are as fairly comparable
as possible measures of topological similarity of edges and nodes,
respectively, and because we are using the same node alignment
strategy in both GREAT and the corresponding node alignment-based
method, any difference that we observe between the two methods will be
a direct consequence of edge-based network alignment compared to
node-based alignment.

Consequently, we denote by NG the node-based alignment method that
uses the greedy node alignment strategy and by NH the node-based
alignment method that uses the Hungarian node alignment strategy. The
node-based alignment method that uses MI-GRAAL's node alignment
strategy is MI-GRAAL itself. Then, we compare each of EGG and EHG to
NG, each of EGH and EHH to NH, and each of EGM and EHM to MI-GRAAL
(Table~\ref{tab:great}).

\begin{figure}[h!]
\centering
\includegraphics[width=1\columnwidth]{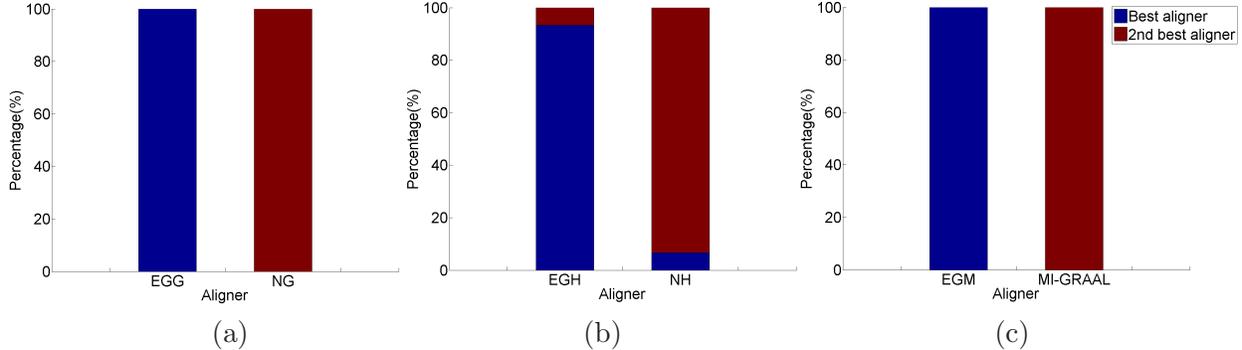}\\
\hspace{0in}(a) \hspace{1.7in}(b) \hspace{1.9in}(c) \\
\vspace{-0.3cm}
\caption{The ranking of a given edge-based network aligner and the
corresponding node-based aligner across all alignments with respect
to all alignment quality measures for: {\bf (a)} greedy, {\bf (b)}
Hungarian, and {\bf (c)} MI-GRAAL's node alignment strategy. Note
that these results are for greedy edge alignment strategy only, but
almost identical results are obtained for optimal edge alignment
strategy, i.e., when comparing EHG with NG, EHH with NH, and EHM
with MI-GRAAL.}
\label{evn}
\end{figure}

Overall, across all network alignment measures, we find that edge
alignment \emph{is} superior to node alignment (Figure~\ref{evn} and
Appendix Figure \ref{evn_detailed}). That is, we demonstrate that
edge-based network alignment outperforms node-based network alignment
when using comparable cost functions and alignment strategies, which
is the main contribution of our study.

\subsection{Speed of greedy edge alignment versus accuracy of optimal
edge alignment}\label{sec:time}

Overall, aligning edges optimally with the Hungarian strategy is
expected to outperform (in terms of accuracy) aligning edges with the
greedy strategy. However, Hungarian method is much slower than the
greedy method, with complexity of $\mathcal{O}(x^3)$ for the former
and $\mathcal{O}(x)$ for the latter, where $x$ is the number of
elements (in this case, edges) to be aligned. So, the question is to
what extent optimal edge alignment improves compared to greedy
alignment, and whether this increase in accuracy is worth the drastic
increase in running time.

\begin{figure}[h]
\centering
\includegraphics[width=1\columnwidth]{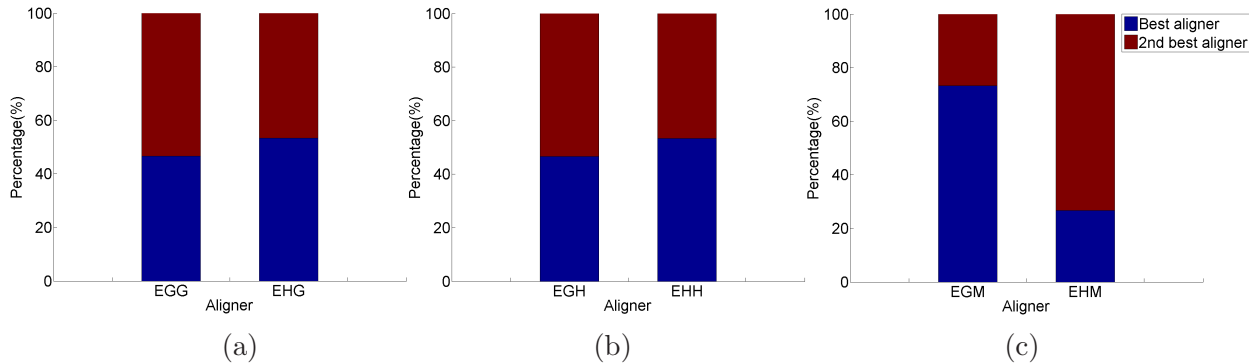}\\
\hspace{0in}(a) \hspace{1.7in}(b) \hspace{1.9in}(c) \\
\vspace{-0.3cm}
\caption{The ranking of a given greedy edge aligner and the
corresponding optimal edge aligner across all alignments with
respect to all alignment quality measures for: {\bf (a)} greedy,
{\bf (b)} Hungarian, and {\bf (c)} MI-GRAAL's node alignment
strategy, for ``within 5\% accuracy''. By this, we mean that the
greedy edge aligner's score is within 5\% of the optimal edge
aligner's score. }
\label{gvh}
\end{figure}

To fairly evaluate this, we compare EGG to EHG, EGH to EHH, to EGM and
EHM. In Table~\ref{tab:time}, we show representative running times and
alignment accuracy scores (in terms of S$^3$ measure) of EGH and EHH,
and in Figure \ref{gvh} we show systematic results for all GREAT
versions while taking into account all measures of alignment quality.
As illustrated, aligning edges with the Hungarian method yields to
only 0\%-14\% increase (depending on the network data) in accuracy
compared aligning edges greedily, but it leads to extremely large
5,271\%-13,407\% increase in running time (Table~\ref{tab:time}).
Further, in the systematic analysis, we find that in 47\%-73\% of all
cases (depending on node alignment strategy) accuracy of greedy edge
alignment is within 5\% of accuracy of optimal edge alignment
(Figure~\ref{gvh}), and the percentages are even higher for being
within 10\% accuracy (Appendix Figure~\ref{gvh_10}). Thus, we believe
that the huge increase in computational complexity of optimal edge
alignment does not justify incremental increase in its accuracy.
Henceforth, especially for large networks, we suggest aligning edges
greedily. (Given the resulting edge alignment-based node cost
function, one still might want to align nodes optimally.)

\begin{table}[h]
\centering
\begin{tabular}{|c|c|c|c|c|c|c|}
\hline
Alignment& \multicolumn{2}{|c|}{EGH} & \multicolumn{2}{|c|}{EHH} & \multicolumn{2}{|l|}{Percentage Increase} \\
\hline
& Time & S$^3$ & Time & S$^3$ & Time & S$^3$ \\
\hline
Yeast-Yeast5\% & 3h 49m 24s & 91.97\% & 220h 20m 03s & 92.33\% &5,271\% & 0.34\% \\
\hline
Yeast-Yeast10\% & 4h 16m 08s & 71.91\% & 432h 29m 14s & 74.05\% &10,031\% & 2.98\%\\
\hline
Yeast-Yeast15\% & 5h 00m 38s & 54.70\% & 570h 48m 22s & 60.13\% &11,292\% & 9.93\%\\
\hline
Yeast-Yeast20\% & 5h 07m 06s & 45.04\% & 691h 18m 40s & 46.77\% &13,407\% & 3.84\% \\
\hline
Yeast-Yeast25\% & 5h 37m 59s & 34.45\% & 755h 36m 06s & 39.25\% &13,314\% & 13.93\% \\ 
\hline
\end{tabular}
\vspace{-0.1cm}
\caption{Computational complexity versus accuracy of greedy 
versus optimal edge alignment. We show the amount of CPU time it took
GREAT variations EGH and EHH to generate an alignment and S$^3$ score
of the resulting alignment, for each pair of yeast networks. We
measure the increase in either running time or accuracy of EHH (i.e.,
optimal edge alignment) over EGH (i.e., greedy edge alignment) as the
difference of the result of EHH and the result of EGH, divided by the
result of EGH. All alignments were ran on the same server with 16
2.3GHz processors and 24GB of RAM.}
\label{tab:time}
\end{table}

\subsection{GREAT versus the most recent and advanced methods}\label{GREAT_vs_existing}

Here, we compare GREAT (the best of all variations) against the
following recent powerful aligners: MI-GRAAL \cite{MIGRAAL}, GHOST
\cite{GHOST}, NETAL \cite{NETAL}, and MAGNA \cite{MAGNA}. We use
default parameters (suggested in the original publications) for all
methods. We also tried SPINAL \cite{SPINAL}, but it did not return
injective node mappings on our network data as the other methods and
was thus excluded.

\begin{figure*}[h]
\includegraphics[width=.41\columnwidth]{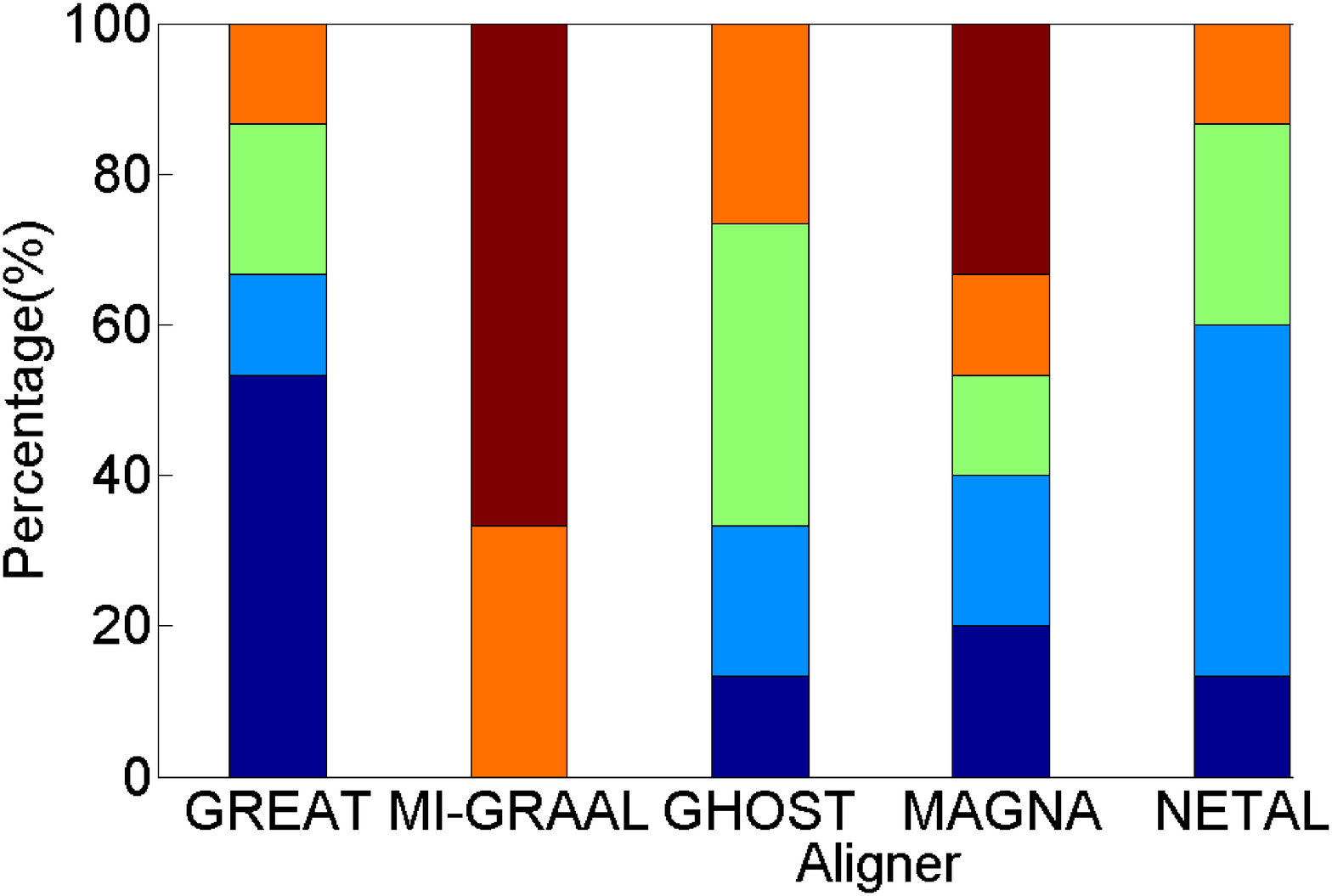}
\includegraphics[width=.41\columnwidth]{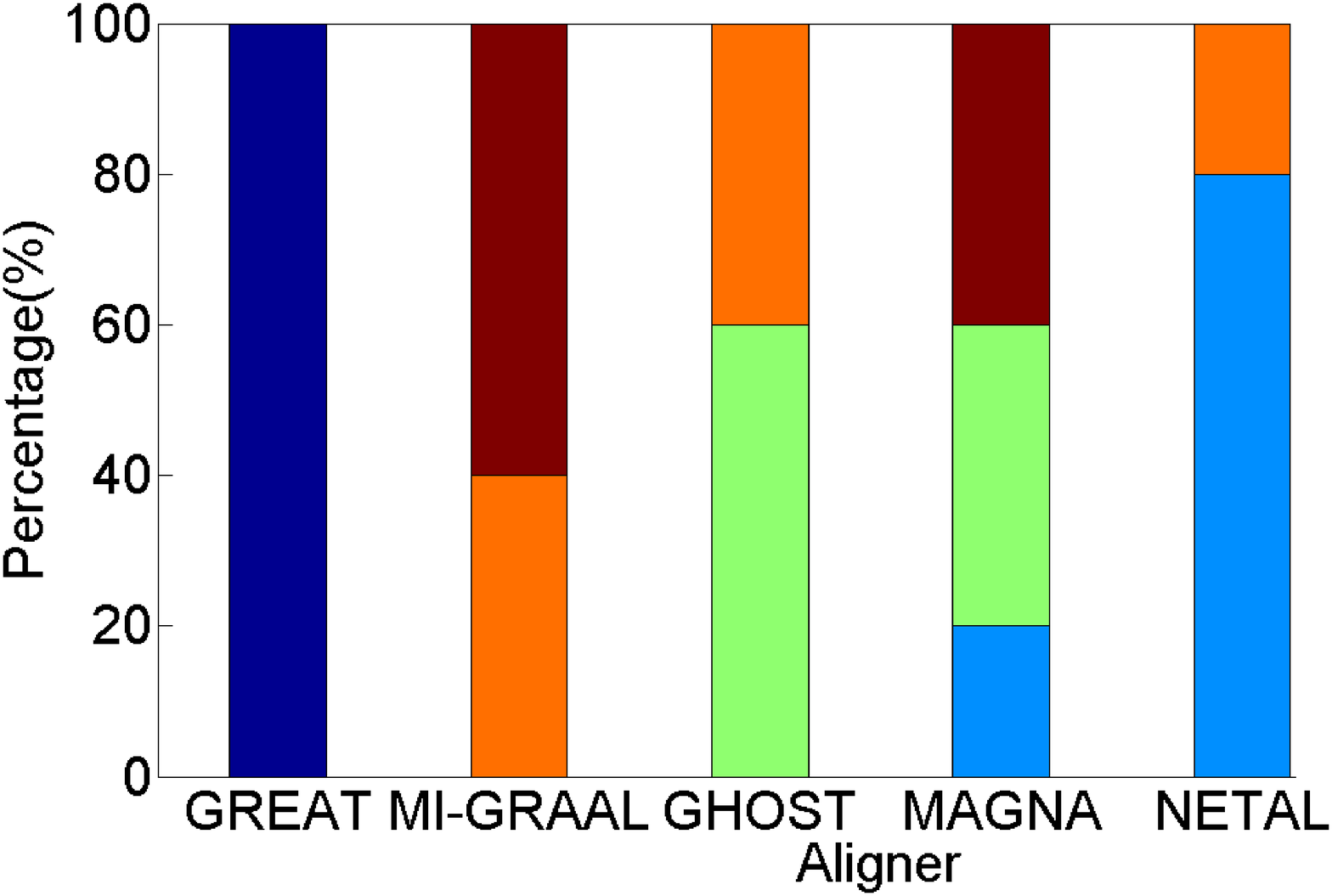}
\includegraphics[width=.15\columnwidth]{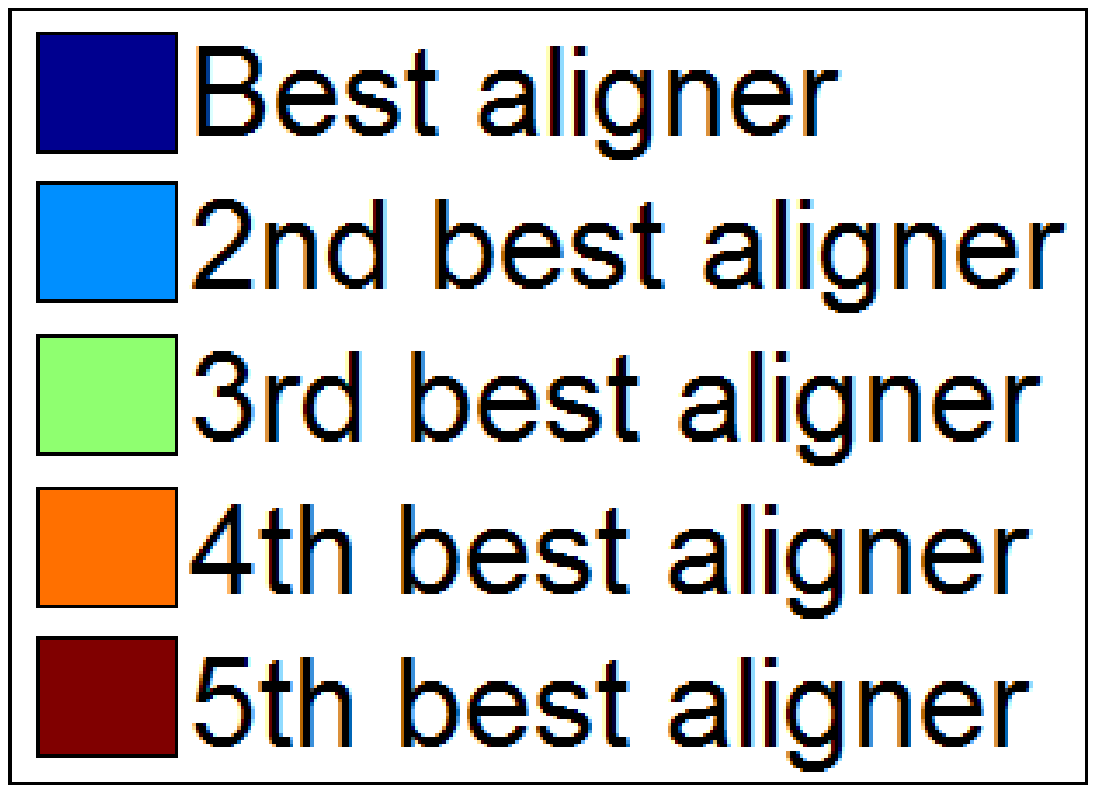}
\hspace{1.4in} {\bf (a)} \hspace{2.4in}{\bf (b)}\\
\includegraphics[width=.41\columnwidth]{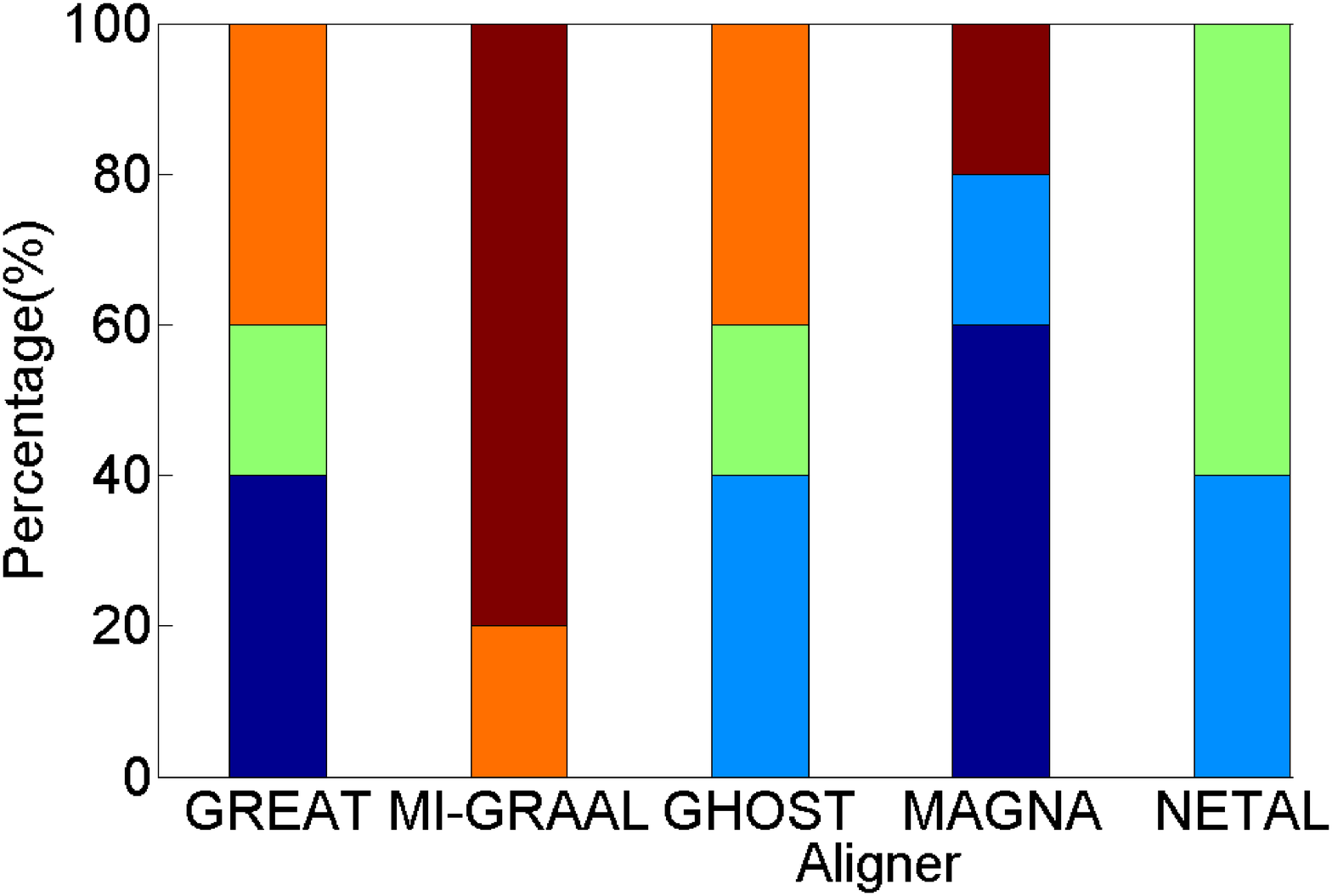}
\includegraphics[width=.41\columnwidth]{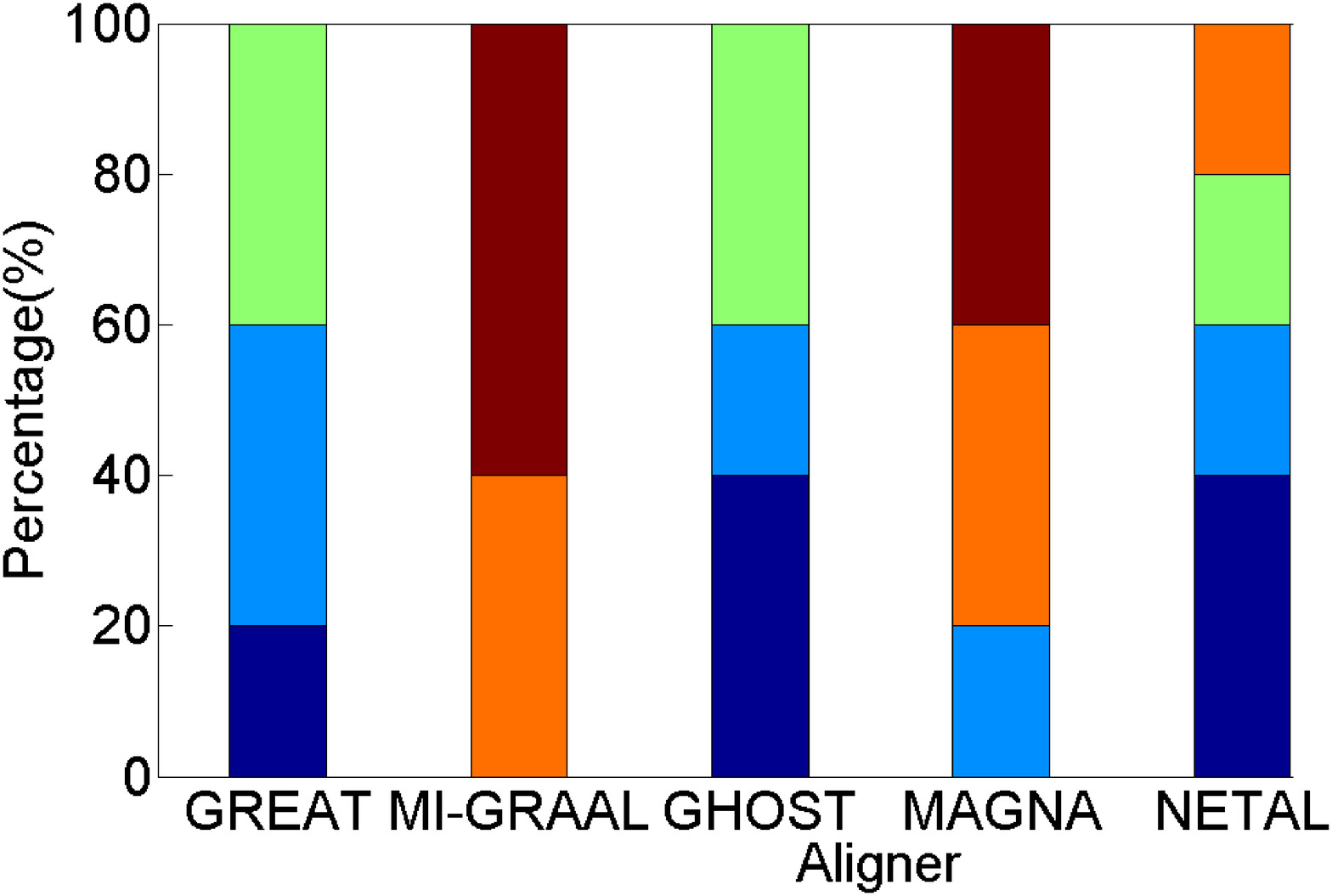}
\includegraphics[width=.15\columnwidth]{legend1.eps}
\hspace{1.4in} {\bf (c)} \hspace{2.4in}{\bf (d)}\\
\vspace{-0.5cm}
\caption{The ranking of GREAT (the best of all variations) and
very recent and thus advanced existing network aligners over all
alignments with respect to: {\bf (a)} all alignment quality measures
combined, {\bf (b)} NC, {\bf (c)} S$^3$, and {\bf (d)} LCCS. }
\label{sums}
\end{figure*}

We find that GREAT is overall the best aligner across all alignment
quality measures (Figure \ref{sums} (a) and Appendix Figure
\ref{yeast}). This is especially true in terms of NC, which is the
ultimate measure of alignment accuracy -- GREAT is \emph{always} the
best of all methods (Figure \ref{sums} (b)). In terms of S$^3$, only
MAGNA is the best ranked in more cases than GREAT (Figure \ref{sums}
(c)). However, this is not surprising, as MAGNA directly optimizes
S$^3$ during alignment construction and is thus expected to dominate
the other methods with respect to this measure. Nonetheless, GREAT
still outperforms MAGNA is 40\% of the cases with respect to S$^3$.
Finally, in terms of LCCS, GREAT is again superior to almost all
methods, including MAGNA (Figure \ref{sums} (d)). Exceptions are
GHOST and NETAL, but GREAT still performs comparably to these methods,
in the sense that all three methods rank as the best or the second
best in equal number (60\%) of all cases.

Thus, we do not only demonstrate that GREAT is superior to fairly
comparable node-based network alignment methods, which differ from
GREAT in a single aspect (edge versus node alignment), but also, it is
superior to the recent state-of-the-art network alignments, which
differ from GREAT in more than one aspect. As such, incorporating
into the design of GREAT the recent methods' algorithmic ideas could
potentially improve GREAT's performance even further.

\vspace{-0.2cm}

\section{Conclusion}

We have presented GREAT, our novel alignment method that aims to
maximize both node and edge conservation, that does so by first
aligning edges well in order to then align nodes well based on their
adjacent edges being aligned well, and that in the process favors
similar conserved edges over dissimilar conserved edges. We have
demonstrated that GREAT, the edge-based network alignment method,
improves upon comparable node-based network alignment methods,
confirming our hypothesis that aligning edges prior to aligning nodes
would improve alignment quality compared to aligning nodes only. In
other words, we have demonstrated superiority of GREAT over methods
that aim to maximize node conservation only, such as MI-GRAAL and
GHOST. At the same time, we have demonstrated superiority of GREAT
over a recent approach that aims to optimize edge conservation only
and that treats each edge the same, namely MAGNA. Finally, we have
shown that GREAT overall outperforms an additional recent
state-of-the-art approach, namely NETAL.

Thus, GREAT (and its modified version that would also account for
functional, e.g., protein sequence, similarities between nodes in
addition to their topological similarities) has important implications
for real-world applications of network alignment to biological
networks of different species, as well as to networks in other
domains, such as social networks or natural language processing. For
example, in computationally biology, GREAT can be used to transfer
aging-related knowledge from well-annotated model species to
poorly-annotated human, thus deepening our current knowledge about
human aging \cite{MilenkovicACMBCB2013,Faisal2014a,Faisal2014}. Or, it
could have implications on user privacy in online social networks, as
network alignment can be used to de-anonymize such network data
\cite{Narayanan2011}.

\vspace{-0.2cm}

\section*{Acknowledgements}
We thank Dr. Horst Bunke for useful discussions regarding GREAT, and
Dr. Seyed Shahriar Arab for his assistance with running NETAL. This work was
supported by the National Science Foundation CCF-1319469 and EAGER
CCF-1243295 grants.

\vspace{0.2cm}

\clearpage

\setcounter{page}{1} \renewcommand{\thepage}{Bibliography Page \arabic{page}}

\bibliographystyle{unsrt}

\small{\bibliography{network_bib} }

\clearpage

\setcounter{page}{1} \renewcommand{\thepage}{Appendix Page \arabic{page}}

\section*{Appendix for:}

\noindent{\large{GREAT: GRaphlet Edge-based network AlignmenT}}\\

\noindent\textbf{Joseph Crawford and Tijana Milenkovi\'{c}$^{*}$}\\ 

\noindent Department of Computer Science and Engineering, Interdisciplinary Center for Network Science and Applications, and ECK Institute for Global Health, University of Notre Dame\\
\noindent$^*$Corresponding Author (E-mail: tmilenko@nd.edu)

\setcounter{figure}{0} \renewcommand{\thefigure}{A.\arabic{figure}}

\begin{figure}[h!]
\centering
\includegraphics[width=.6\columnwidth]{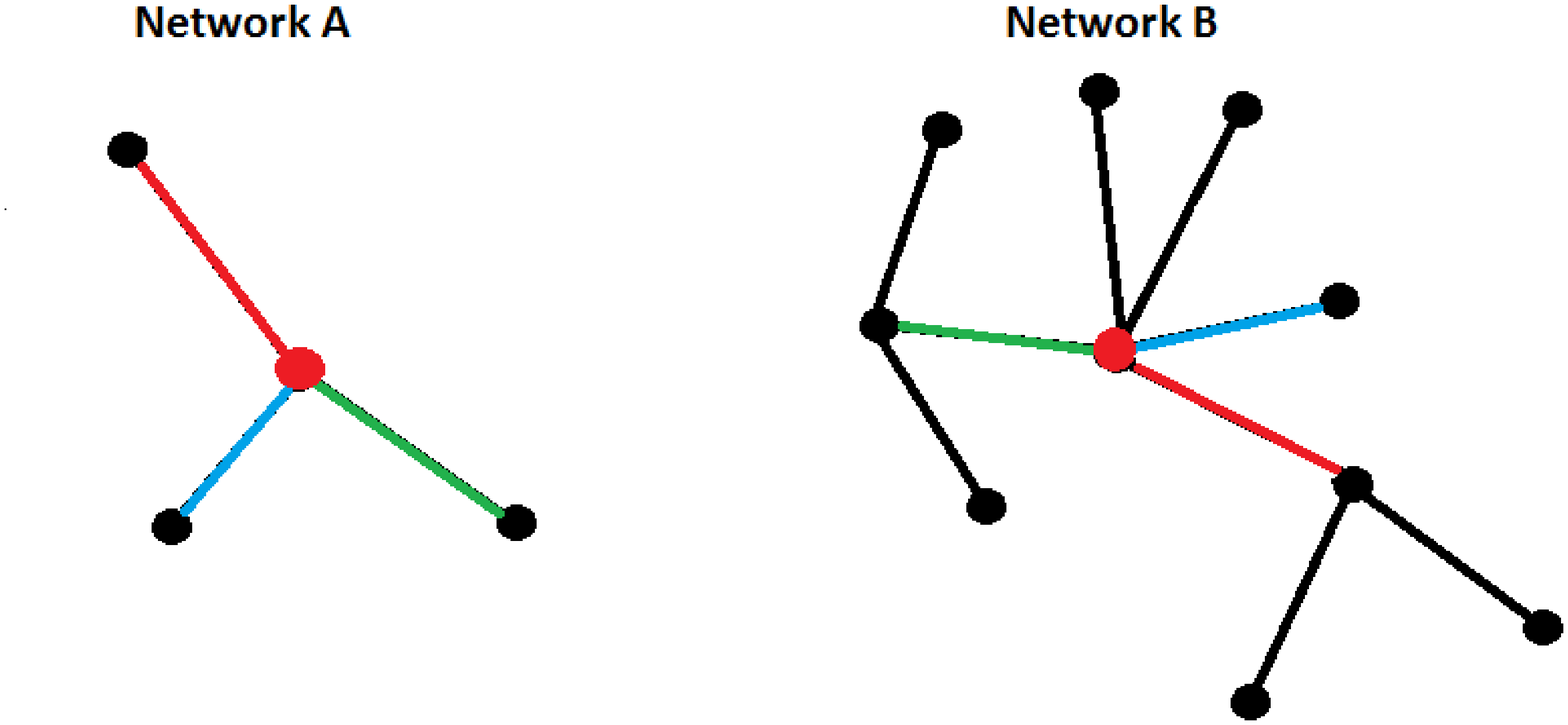}\\
\vspace{-0.3cm}
\caption{Illustration of GREAT's edge alignment-based node cost
function. To measure similarity between two red nodes in networks A
and B, GREAT first identifies all edges adjacent to the red nodes that
are aligned to each other, along with their similarity scores with
respect to edge cost function. In this case, let us assume that red
edge in A is aligned to red edge in B with score of 0.9, blue edge in
A is aligned to blue edge in B with score of 0.8, and green edge in A
is aligned to green edge in B with score of 0.7, while all black edges
in the larger of the two networks, i.e., network B, are unaligned.
Then, GREAT's edge alignment-based node cost function is the sum of
similarities (with respect to edge cost function) over all aligned
edges.}
\label{ncf}
\end{figure}

\begin{figure}[h]
\centering
\includegraphics[width=1\columnwidth]{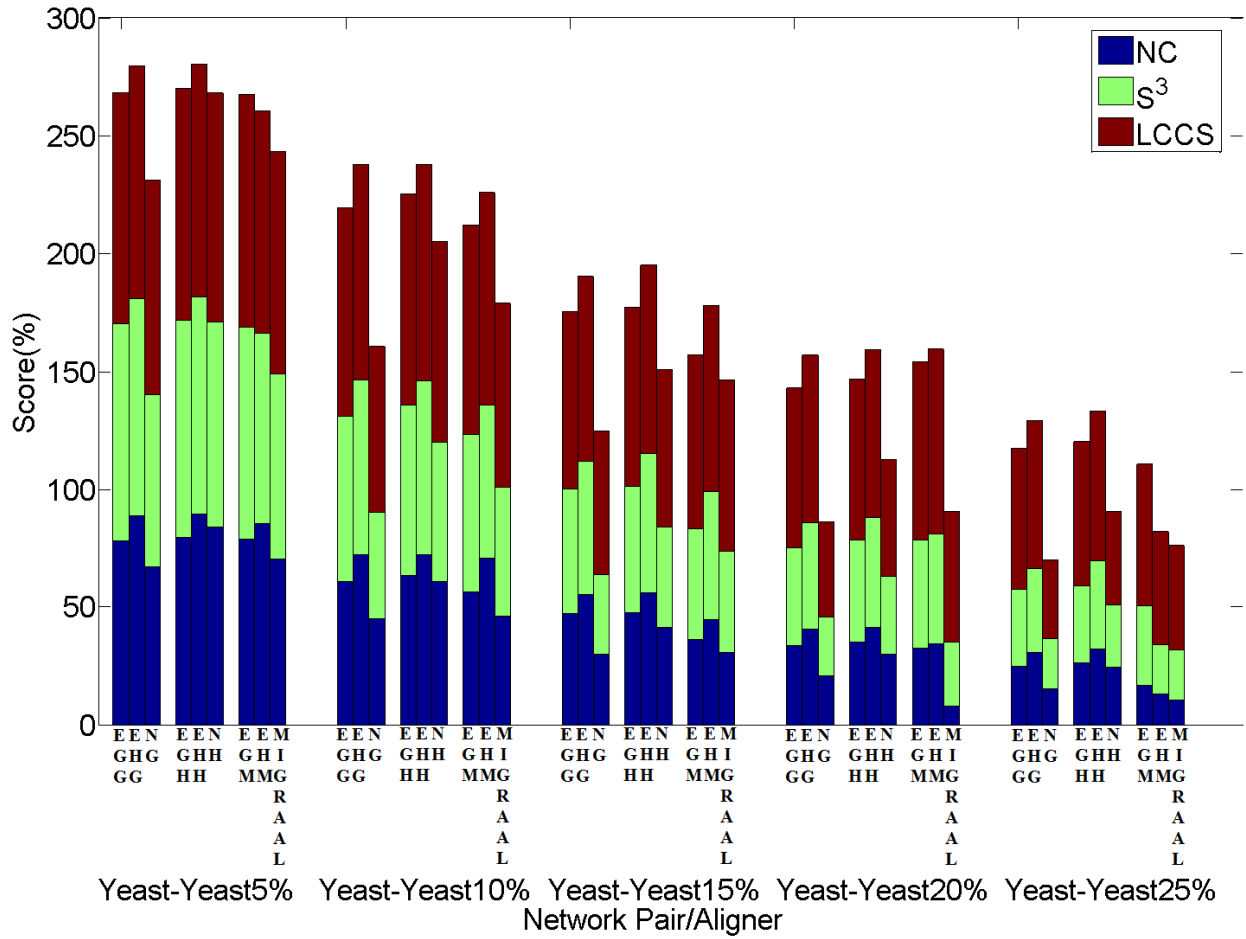}\\
\caption{Alignment quality results of all six variations of GREAT edge-based 
network alignment method and the corresponding three node-based
network alignment methods, for each of the five network pairs and with
respect to each of the three alignment quality measures. }
\label{evn_detailed}
\end{figure}

\begin{figure}[h]
\centering
\includegraphics[width=1\columnwidth]{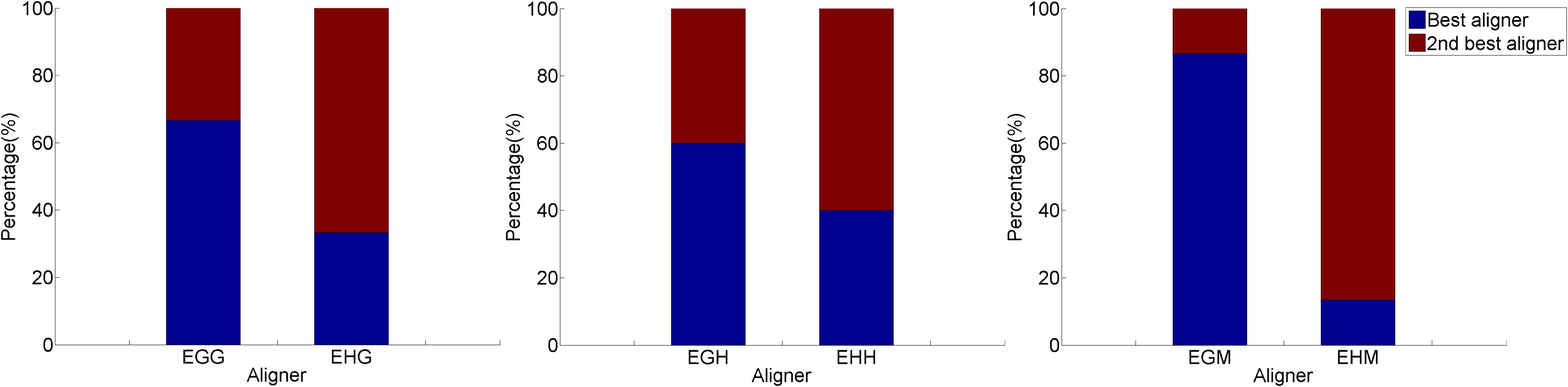}\\
\hspace{0in}(a) \hspace{1.7in}(b) \hspace{1.9in}(c) \\
\caption{The ranking of a given greedy edge aligner and the
corresponding optimal edge aligner across all alignments with
respect to all alignment quality measures for: {\bf (a)} greedy,
{\bf (b)} Hungarian, and {\bf (c)} MI-GRAAL's node alignment
strategy, for ``within 10\% accuracy''. By this, we mean that the
greedy edge aligner's score is within 10\% of the optimal edge
aligner's score. }
\label{gvh_10}
\end{figure}

\begin{figure}[h]
\centering
\includegraphics[width=1\columnwidth]{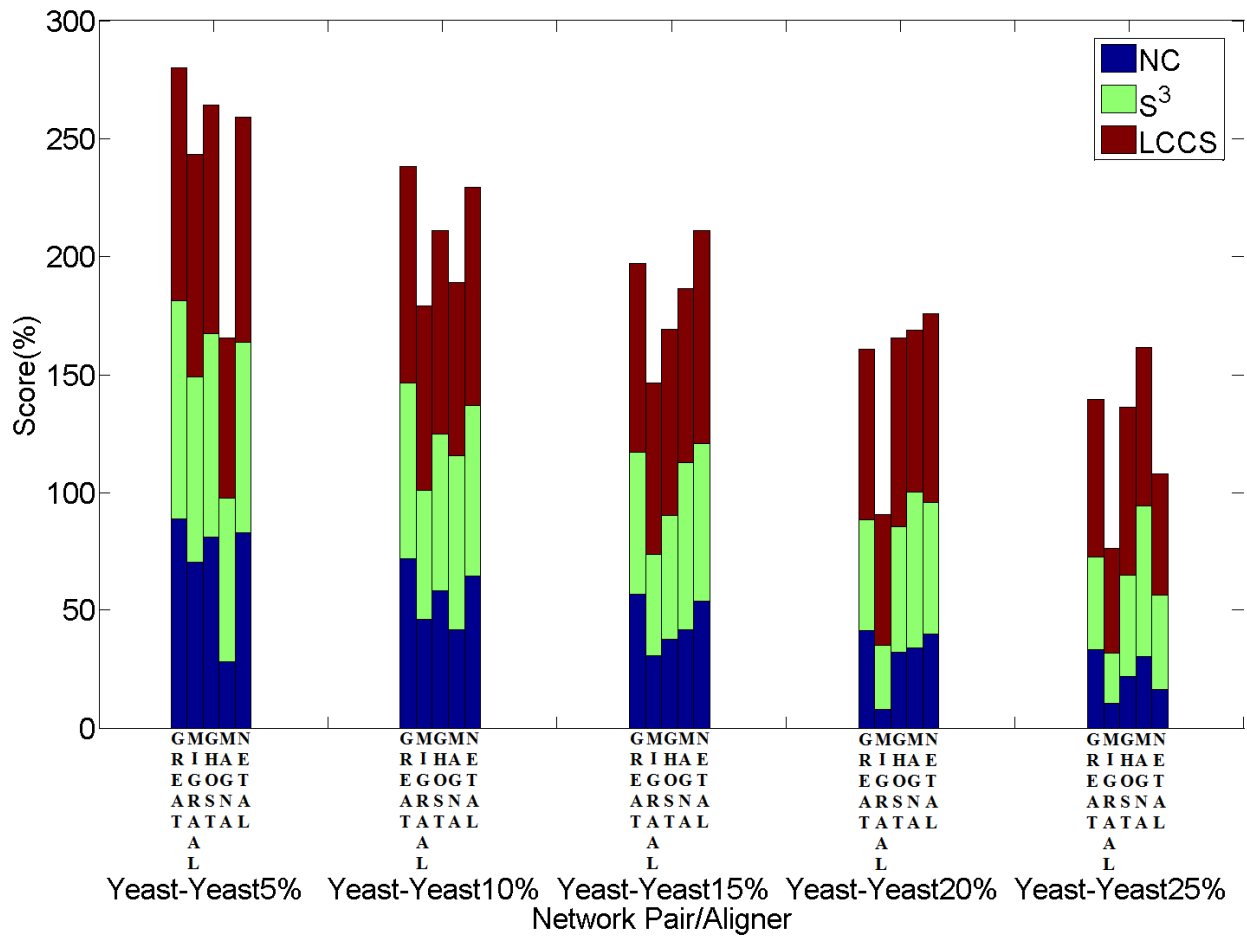}\\
\caption{Alignment quality results of GREAT (the best of all variations) and 
four very recent and thus advanced existing network aligners, for each
of the five network pairs and with respect to each alignment quality
measure. }
\label{yeast}
\end{figure}

\end{document}